\documentclass[aps,prd,a4paper,twocolumn,showpacs,nofootinbib]{revtex4}
\usepackage{amssymb,amsmath,latexsym,mathrsfs}
\usepackage{graphicx,subfigure}
\usepackage{epsfig}
\usepackage{varioref,xr-hyper}
\usepackage{color}

\usepackage[hypertex,draft=false,verbose=false,debug=false
            hyperindex=true,hypertexnames=true]{hyperref}
\begin{document}

\title{Varying couplings in the early universe: correlated variations of $\alpha$ and $G$}

\author{C.J.A.P. Martins$^{a,b}$}
\author{Eloisa Menegoni$^{c}$}
\author{Silvia Galli$^{d,e}$}
\author{Gianpiero Mangano$^{f}$}
\author{Alessandro Melchiorri$^{d}$}

\affiliation{$^a$ Centro de Astrof\'{\i}sica, Universidade do Porto, Rua das Estrelas, 4150-762 Porto, Portugal}
\affiliation{$^b$ CTC, DAMTP, University of Cambridge, Wilberforce Road, Cambridge CB3 0WA, United Kingdom}
\affiliation{$^c$ ICRA, International Center for Relativistic Astrophysics, University of Rome, ``La Sapienza,'' P.le Aldo Moro 2, 00185 Rome, Italy}
\affiliation{$^d$ Physics Department and INFN, Universit\`a di Roma ``La Sapienza'', Ple Aldo Moro 2, 00185 Rome, Italy}
\affiliation{$^e$ Laboratoire Astroparticule et Cosmologie (APC), Universit\'e Paris Diderot, 75205 Paris cedex 13}
\affiliation{$^f$  INFN and Physics Department, University of Naples ``Federico II'', Via Cintia, 80126 Naples, Italy}

\begin{abstract}
The Cosmic Microwave Background anisotropies provide a unique opportunity to constrain simultaneous variations of the fine-structure constant $\alpha$ and Newton's gravitational constant $G$. Those correlated variations are possible in a wide class of theoretical models. In this brief paper we show that the current data, assuming that particle masses are constant, gives no clear indication for such variations, but already prefers that any relative variations in $\alpha$ should be of the same sign of those of $G$ for variations of $\sim 1 \%$. We also show that a cosmic complementarity is present with Big Bang Nucleosynthesis and that a combination of current CMB and BBN data strongly constraints simultaneous variations in $\alpha$ and $G$. We finally discuss the future bounds achievable by the Planck satellite mission.
\end{abstract}
\pacs{98.80.Cq, 04.50.+h, 95.35.+d, 98.70.Vc}

\maketitle

\section{Introduction}

The behavior of nature's fundamental couplings is a subject of much recent interest. There is ample experimental evidence showing that they run with energy, and many particle physics and cosmology models suggest that they should also roll with time. This explains why the European Space Agency (ESA) and the European Southern Observatory (ESO) now list varying fundamental constants among their key science drivers for the next generation of facilities. Recent technological developments have for the first time provided us with tools to accurately test this hypothesis. Two recent reviews \cite{uzan,GarciaBerro} discuss most of these developments.

In particular, there are controversial claims for time variations of the fine-structure constant $\alpha$ \cite{Murphy} and the proton-to-electron mass ratio $\mu$ \cite{Reinhold} at redshifts $z\sim1-3$, as well as of possible spatial variations of the latter in the Galaxy \cite{Molaro}. While these have so far not been confirmed by independent analyses \cite{Srianand,Levshakov,King,Thompson}, their potential implications certainly warrant further study. In the present work we focus on the cosmic microwave background (CMB) as a means to probe the behavior of these couplings at redsifts $z\sim10^3$, but we will also discuss the importance of constraints at much higher redshifts ($z\sim10^{10}$), coming from Big Bang Nucleosynthesis.

Typically, in any sensible theory where a coupling is rolling, one generically expects the others to do so as well, though possibly at fairly different rates. Since the rolling is expected to be due to the same underlying mechanism (the most natural of which will be a dynamical, fundamental scalar field), the rates of change of the various couplings will in fact be related in any given theory.

From an experimental point of view, one can either take the simplifying assumption that only the coupling one wants to constrain is varying while the others are constant, or try to constrain joint variations at the expense of a more complicated analysis. In the latter case one can either choose a particular set of such relations (thereby constraining only a particular theory, though usually quite tightly) or phenomenologically treat the different variations as independent (thus obtaining model-independent but usually weaker constraints). In this work we will use the CMB to constrain possible variations of $\alpha$ and Newton's constant $G$, assuming them to be related by a phenomenological parameter whose value will be different in various fundamental physics scenarios.

The importance of $\alpha$ as a fundamental physics probe stems from the fact that it is ubiquitous in electromagnetic processes. In the past this has been extensively used to constrain the fine-structure constant \cite{hannestad,kaplinghat,avelino,battye,rocha,ichikawa,petruta,jap}. These studies yielded results consistent with no variation, but due to degeneracies with other cosmological parameters the accuracy is lower than that of low-redshift measurements: it is only with the latest available data that constraints stronger than the percent level have been obtained \cite{Menegoni}.

Speaking of variations of dimensional constants obviously has no physical significance: one can design any variation by defining appropriate units of length, time and energy. However, one is free to choose an arbitrary dimensionful unit as a standard and compare it with other quantities. If one assumes particle masses to be constant (as we will do in this paper), constraints on the gravitational constant $G$ are in fact constraining the dimensionless product of $G$ and the nucleon mass squared. With this caveat, constraints on a rolling $G$ provide key information on the gravitational sector. Somewhat paradoxically, $G$ was the first constant to be measured but is now the least well known, a result of the weakness of gravity. Indeed, in the past two decades our knowledge of its value didn't substantially improve from the precision of $0.05 \%$ reached in $1942$ (see \cite{1942}). Recent laboratory measurements (see e.g. \cite{gother}) point towards an uncertainty at the level of $\sim 0.4 \%$, while other works claim an improved precisions below $0.01 \%$ (\cite{claims}). Analysis of the secular variation of the period of nonradial pulsations of the white dwarf G117-B15A (\cite{gwd}) has produced complementary constraints at $\sim 0.1 \%$ level.

On the other hand, as shown in \cite{zalzahn} a variation in the gravitational constant could also affect the CMB anisotropy spectra, with current bounds of the order of $\sim 10 \%$ (se e.g. \cite{galli2}) and constraints at level of $\sim 1 \%$ achievable with future CMB experiments as Planck.

The interesting point is that the CMB is an observable potentially sensitive to variations in {\it both} fundamental constants. It is therefore timely to 
perform a combined analysis of CMB data considering simultaneous variations in $\alpha$ and $G$ in order to investigate the possible correlations and
deviations from the standard values.

Specifically, we will consider that the variations of $\alpha$ and $G$ are related by
\begin{equation}
\frac{\Delta\alpha}{\alpha} = Q \frac{\Delta G}{G}
\end{equation}
with $Q$ a free parameter that can be positive or negative, but not much larger than unity in absolute value (we will conservatively assume that $-10 < Q < 10$). As an illustration of the range of values allowed in some representative models, Kaluza-Klein-type theories typically have $1<Q\le3$, Einstein-Yang-Mills has $Q=1$, and Randall-Sundrum type models have very small positive $Q$s (say $Q\sim~0.01$). These examples are discussed in more detail in \cite{GarciaBerro,Loren}---note that all of them have $Q>0$. However, one can equally easily find models with $Q<0$: for example string theory dilaton-type models have $Q\sim -1$ \cite{Damour}, while the BSBM-Brans-Dicke model has $Q=-1$ exactly \cite{Barrow}. 

In the rest of the paper we briefly sketch out analysis pipeline in Sec. \ref{sec2} and present our results in Sec. \ref{sec3}. Finally we will discuss the implications of our results and present some conclusions. We emphasize that the key assumption of the present analysis is that the particle masses are kept constant, and there are no changes to the strong sector. A more general analysis, without this assumption, will be presented elsewhere.

\section{Analysis Method\label{sec2}}

We allow for a possible variation in the fine structure constant and
in the Newton's constant during recombination using the method described 
in \cite{avelino} and \cite{zalzahn},
modifying the publicly available RECFAST (\cite{recfast}) routine
in the CAMB (\cite{camb}) CMB code.
As in \cite{zalzahn} we consider variations in the Newton's constant $G$
by introducing a new dimensionless parameter $\lambda_G$ such that

\begin{equation}
G \Rightarrow \lambda_G^2 G
\end{equation}

\noindent while we constrain variations in the fine structure constant by considering 
the dimensionless parameter $\alpha / \alpha_0$ where $\alpha_0$ is the
fine structure constant today.

The analysis method we adopt is based on the
publicly available Markov Chain Monte Carlo package \texttt{cosmomc}
\cite{Lewis:2002ah} with a convergence diagnostics done through the Gelman and Rubin statistics.

We sample the following ten-dimensional set of cosmological parameters, adopting flat priors
on each of them: the baryon and cold dark matter densities $\omega_{\rm b}$ and
$\omega_{\rm c}$, the Hubble constant $H_0$, the scalar spectral index $n_s$,
the overall normalization of the spectrum $A_s$ at $k=0.05$ Mpc$^{-1}$,
the optical depth to reionization, $\tau$ and, finally, the variations in the
fine structure constant $\alpha / \alpha_0$ and in the Newton's constant $\lambda_G$.
Furthermore, we consider purely adiabatic initial conditions and we impose spatial flatness.

Our basic data set is the five--year WMAP data \cite{wmap5cosm,wmap5komatsu}
(temperature and polarization) with the routine for computing the likelihood supplied by the WMAP team.
In addition to the WMAP data we also consider the following CMB datasets:
ACBAR (\cite{acbar}), QUAD (\cite{quad}) and BICEP (\cite{bicep}), as well as
the older datasets from BOOMERanG (\cite{boom03}) and CBI (\cite{cbi}).
For all these experiments we marginalize over a possible contamination
from Sunyaev-Zeldovich component, rescaling the WMAP template at the corresponding
experimental frequencies. 

In what follows, we also combine the CMB data with the recent UNION catalog of
supernovae type Ia luminosity distances and with the improved constraint on the Hubble 
constant of $h=0.747\pm0.036$ at $68 \%$ c.l.. from the recent analysis of \cite{riess}.

Constraints on $\lambda_G$ and $\alpha$ are also computed using
standard BBN theoretical predictions as provided by the 
numerical code described in \cite{Pisanti:2007hk}, which includes
a full updating of all rates entering the
nuclear chain based on the most recent experimental results on
nuclear cross sections. The BBN predictions are compared with the experimental determinations of
the $^4$He mass fraction $Y_p$ and D/H abundance ratio, as discussed in \cite{Serpico:2004gx}
\begin{eqnarray}
Y_p &=& 0.250 \pm 0.003 \\
\textrm{D/H}&=&(2.87_{-0.21}^{+0.22}) \cdot 10^{-5} \label{deut}
\end{eqnarray}

\section{Results\label{sec3}}

\begin{figure}[h!]
\includegraphics[width=7.2cm]{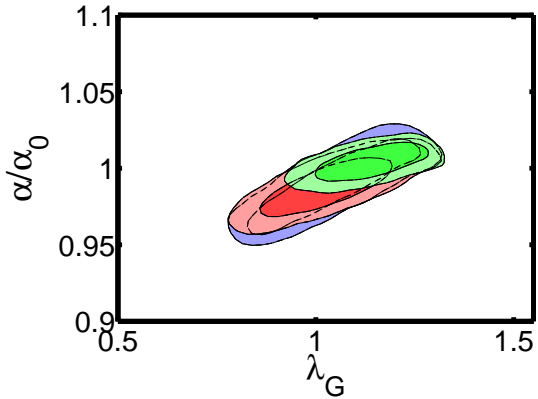}
\caption{\label{plot1}$68 \%$ and $95 \%$ c.l. constraints on the $\alpha / \alpha_0$ vs
$\lambda_G$ for different datasets. The contours regions come from CMB data (blue),
CMB data and SN-Ia (red), and CMB+HST (green).}
\end{figure}

In Table \ref{tab:wdm} we report the constraints on the $\alpha /\alpha_0$ 
and the $\lambda_G$ parameters obtained from the COSMOMC analysis, using the
the different combinations of the datasets described in the previous section,
and in Figure \ref{plot1} we show the $68 \%$ and $95 \%$ c.l. constraints on the $\alpha / \alpha_0$ vs
$\lambda_G$ for the different datasets.

\begin{table}[h!]
\begin{center}
\begin{tabular}{lclllll}
Experiment & & $\alpha / \alpha_0$ & $68\%$ c.l.& $\lambda_G$ & $68\%$ c.l. \\
\hline
\vspace{0.2cm}
All CMB & & $0.999$ & $\pm 0.017$ & $1.04$ & $\pm 0.12$ \\
\vspace{0.2cm}
All CMB+SN-Ia & & $0.989$ & $\pm 0.012$ & $1.04$ & $\pm 0.11$\\
\vspace{0.2cm}
All CMB+HST & & $1.003$ & $\pm 0.008$ & $1.13$ & $\pm 0.09$\\
\vspace{0.2cm}
ALL CMB+BBN && $0.985$& $\pm 0.009$ & $1.01$ & $\pm 0.01$\\
\vspace{0.2cm}
Planck only && $1.000$& $\pm 0.015$ & $1.02$ & $\pm 0.09$\\
\hline
\end{tabular}
\caption{Limits on $\alpha / \alpha_0$ and $\lambda_G$ from CMB data only (first row), from CMB+SN-Ia (second row),
from CMB plus the HST prior on the Hubble constant, $h=0.748\pm0.036$ (third row), from CMB plus BBN (fourth row)
and for simulated mock data for the Planck experiment. We report errors at $68\%$ confidence level.}
\label{tab:wdm}
\end{center}
\end{table}

Comparing these with the results of the recent studies for the two individual parameters \cite{galli2,Menegoni}
we see only a mild changes in the best fit and confidence intervals for $\alpha$, while the changes are
somewhat larger for $G$.

\begin{figure}[h!]
\includegraphics[width=7.2cm]{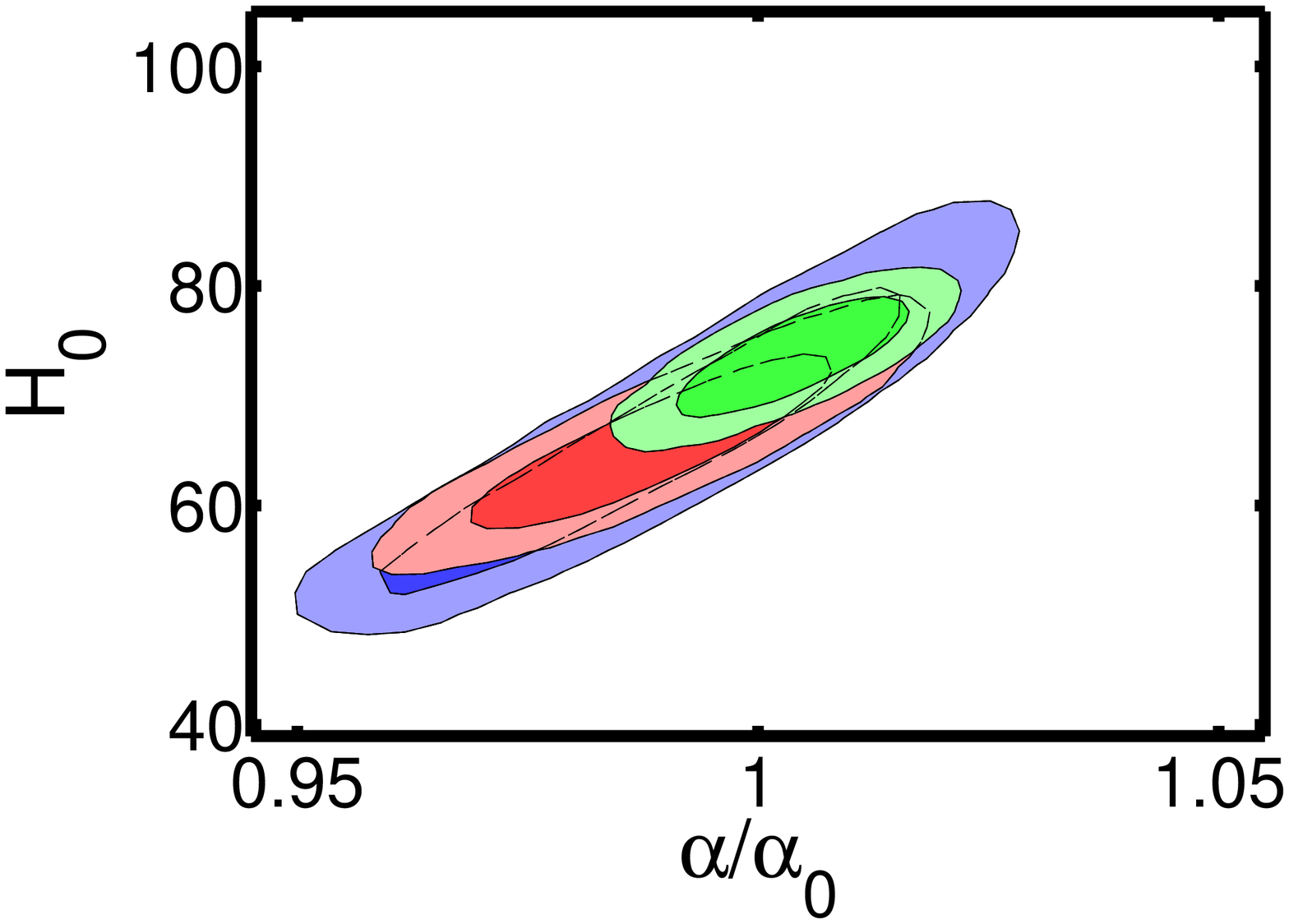}
\caption{\label{plot2}$68 \%$ and $95 \%$ c.l. constraints on the $\alpha / \alpha_0$ vs
$H_0$ for different datasets. The contours regions come from CMB data (blue),
CMB data and SN-Ia (red), and CMB+HST (green).}
\end{figure}

\begin{figure}[h!]
\includegraphics[width=7.2cm]{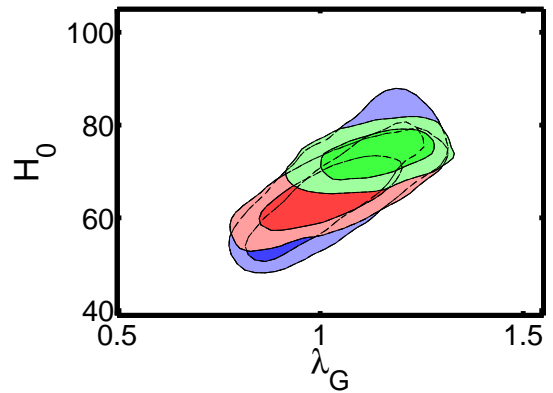}
\caption{\label{plot3}$68 \%$ and $95 \%$ c.l. constraints on the 
$\lambda_G$ vs $H_0$ for different datasets. The contours regions come from CMB data (blue),
CMB data and SN-Ia (red), and CMB+HST (green).}
\end{figure}

More importantly, a degeneracy is clearly present between $\lambda_G$
and the fine structure constant. The underlying reason is easy to understand.
A change in $\alpha$ shifts the recombination epoch, affecting the
angular diameter distance at recombination and the peaks position in the CMB anisotropy angular spectra.
A similar effect can be obtained by changing the value of $\lambda_G$ and the two parameters are
therefore degenerate. Both parameters are degenerate with the Hubble constant $H_0$ as we can see
from Figures \ref{plot2} and \ref{plot3}.

From the above it follows that including the recent HST measurements of $H_0$ has an important effect: it
breaks the $\alpha$-$H_0$ and $\lambda_G$-$H_0$ degeneracies and thereby provides a stronger bound on those parameters. 
This is clearly shown in the third row of Table 1.

As already mentioned in the introduction, from the perspective of mode-building it is interesting to consider the behavior of the $Q$ parameter, 
which in terms of our analysis pipeline is defined as

\begin{equation}
\frac{\alpha}{\alpha_0}-1=Q(\lambda_G^2-1).
\end{equation}

Obviously, since the current data doesn't show any evidence for variations in $G$ or $\alpha$ it is not possible to rule out
any of the models discussed before. The CMB data, however, does show a clear correlation between $\alpha$ and $\lambda_G$ that 
could be fitted with a linear relation. Considering the models that are inside the $95 \%$ confidence level 
we found constraints on the best-fit $Q$ parameter as reported in Table \ref{tab:q}.

\begin{table}[h!]
\begin{center}
\begin{tabular}{lclll}
Dataset & & Constraint on $Q$ (at $68\%$ c.l.) \\
\hline
\vspace{0.2cm}
All CMB & & $0.844 < Q < 0.888$  \\
\vspace{0.2cm}
All CMB+SN-Ia & & $0.868 < Q < 0.912$\\
\vspace{0.2cm}
All CMB+ HST & & $0.744 < Q < 0.780$\\
\vspace{0.2cm}
Planck only & & $0.872 < Q < 0.900$\\

\hline
\end{tabular}
\caption{Limits on the $Q$ parameter from CMB data only (first row), from CMB+SN-Ia (second row),
and from CMB plus the HST prior on the Hubble constant, $h=0.748\pm0.036$ (third row).
We report errors at $68\%$ confidence level. This analysis included only the models that are inside the $95 \%$ confidence level.}
\label{tab:q}
\end{center}
\end{table}

The data therefore prefers a value of $Q$ whose value is significantly larger than zero, and again we see that the HST prior has a very strong effect. Taken at face value, this result would therefore strongly disfavor the models we mentioned in the introduction as having $Q$ negative for $\lambda_G\sim0.1$ in absolute value. In other words, if there are any variations at or around the percent level, then the variations of the two parameters must have the same sign.

\begin{figure}[h!]
\includegraphics[width=7.2cm]{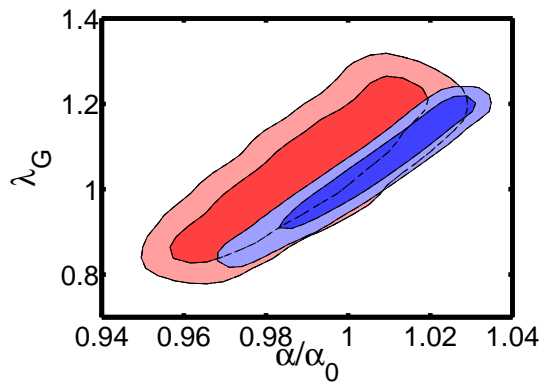}
\caption{\label{plot4}$68 \%$ and $95 \%$ c.l. constraints on the 
$\lambda_G$ vs $\alpha/\alpha_0$ expected from the Planck satellite alone (Blue) compared with the constraints 
obtained from current CMB data (Red). The degeneracy between these two parameters doesn't allow for significant improvements
respect to current bounds.}
\end{figure}

It is also interesting to notice that, due to the degeneracy between the two parameters, a future detection for a variation in $\alpha$ could be
on the contrary due to a variation in $G$. It is therefore important to pursue a combined search for variations in the two constants since
their effect on the CMB anisotropy are very similar. In this respect we forecast the constraints achievable from the Planck satellite mission by using the specifications described in \cite{galli,galli2}, assuming a standard $\Lambda$-CDM model. These results are also presented in Tables I and II. As we can see, due to the $\alpha$-$G$ degeneracy the bounds obtained by Planck will be only marginally improved respect from current CMB data, as also shown in Figure \ref{plot4}.

However, we again emphasize that these results were obtained on the assumption that $\alpha$ and $G$ are the only rolling couplings, while everything else is kept fixed. While some of the above models should be taken as toy models and have little to say about this assumption, there are certainly others where one does expect particle masses to vary. Among other effects, this will lead to variations of $\mu$ which may again have an imprint on the CMB, so a full analysis needs to be done before any stronger statements can be made. One analysis along these lines has been recently carried out \cite{Nakashima}, but the simplifying assumptions made by the authors imply that it applies only to a specific class of models, so a general analysis is still missing.

Moreover, our analysis is made under the approximation of $\alpha$ and $G$ as constant in time and space. While the CMB constraints come mainly from recombination epoch and on scales larger than $10 Mpc$, it is possible that our constraints could vary when a more accurate evolution in time and space up to the current epoch 
is considered.

If we assume that $\alpha$ and $G$ do not vary from BBN to recombination we can combine the CMB results with a BBN analysis. (This is a reasonable assumption, since in many scalar-field based models the field is frozen during the radiation era.) 
The results are plotted in Figure $5$ and also reported in Table 1. 

As we can see a sort of cosmic complementarity is present between the two datasets and much stronger bounds can be achieved. This can be understood as follows. Differently than for the CMB, in the case of BBN in fact, variations of $\alpha$ and $G$ are negatively correlated, since both $Y_p$ and Deuterium are increasing functions of both parameters, see e.g. \cite{Serpico:2004gx,Dent:2007zu}. This implies that the likelihood contours for BBN and CMB are almost orthogonal in the $\alpha-\lambda_G$ plane, thus leading to a tighter bound, in particular on $\lambda_G$.

\begin{figure}[h!]
\includegraphics[width=7.2cm]{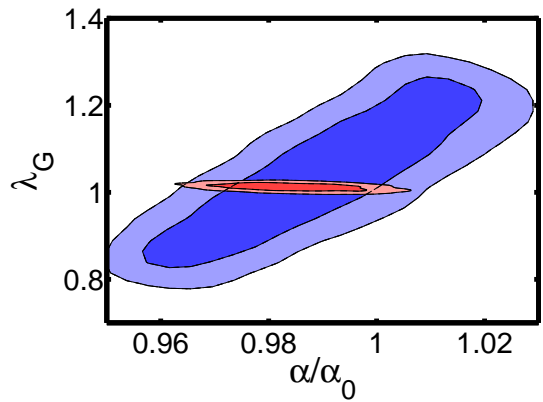}
\caption{\label{plot5}$68 \%$ and $95 \%$ c.l. constraints on the 
$\lambda_G$ vs $\alpha/\alpha_0$ obtained from current CMB data (Blue) compared with the constraints 
obtained from current CMB data plus BBN bounds (Red). A cosmic complementarity between those two datasets helps in breaking the degeneracy and
a much stronger bound is achievable on $\lambda_G$.}
\end{figure}

\section{Discussion and Conclusions}

In this paper we have studied the effects of simultaneous variations of the fine-structure constant $\alpha$ and Newton's gravitational constant $G$ on the CMB. Our results indicate that the current data gives no clear indication about the relative sign of the variations, but already prefers that any relative variations in $\alpha$ should be of the same sign of $G$ for $\sim 1 \%$ variations. We have also shown that, under the assumption that there's no rolling during the radiation era, much tighter constraints can be obtained by adding BBN data.

While our analysis is more general (and more robust) than previous studies that have considered only the variation of one of these couplings, it is by no means the final word on the subject. As a convenient simplification, we have assumed all particle masses to be constant. This assumption is often made (implicitly, if nothing else) in varying $G$ studies since it gives an unambiguous meaning to its variations, but it is not a natural one. In particular, one expects that in any sensible model where there are $\alpha$ variations, other quantities such as the proton-to-electron mass ratio $\mu$ will also vary. A more general analysis, allowing for this possibility, is obviously much more difficult to carry out in full generality, but nevertheless still feasible. One such analysis will be presented in a follow-up paper.

In any case, our results already show how fairly standard astrophysical observables can place strong constraints on high-energy physics models that would otherwise be difficult to test in laboratory or accelerator settings. With the significant gains in sensitivity expected for the next generation of ground and space experiments these constraints will become much stronger, and the early universe will become a key laboratory in which to probe fundamental physics.

\begin{acknowledgments}

The work of C.M. is funded by a Ci\^encia2007 Research Contract, funded by FCT/MCTES (Portugal) and POPH/FSE (EC).

\end{acknowledgments}

\end{document}